# Modeling and Simulation of Passenger Traffic in a National Airport

J. Enciso, M.Sc.[1], J. Vargas, Ing.(c)[1], and P. Martinez, Ing.[1]
[1]Universidad de los Llanos, Colombia, jenciso@unillanos.edu.co, juandavidvargasmora@gmail.com, pablo.martinez7463@gmail.com

*Abstract*– Optimal operation of a country's air transport infrastructure plays a major role in the economic development of nations. Due to the increasing use of air transportation in today's world, flights' boarding times have become a concern for both airlines and airports, thus the importance of knowing beforehand how changes in flights demand parameters and physical airport layout will affect passengers flow and boarding times. This paper presents a pedestrian modeling study in which a national airport passenger flow was analyzed. The study was conducted at Vanguardia National Airport in Villavicencio, Meta, Colombia. Different effects of structural changes are shown and provide judging elements for decision makers regarding passenger traffic in airport design.

## I. INTRODUCTION

The aim of this research project is to understand the passenger flow in a national airport using a pedestrian model with emphasis on boarding time (the time it takes for a passenger to go from the airport's entrance door to the airplane, going trough check-in and control points, and waiting halls) in order to provide insights for decision makers regarding the capacity usage and passenger satisfaction trough diminishing waiting times.

Technological advancements have made flights safer, faster, and cheaper, making air transportation an each time more popular option for passengers. Passenger satisfaction, in accordance with cost and operational efficiency, is increasingly difficult to manage due to growing demand, thereby, increasing demand becomes a double-edged sword: while it may seems like a perfect opportunity to increase profits for airports and airlines, it can also become a problem as the rising demand makes the boarding and landing times longer, that is, each flight needs more time and in consequence, a smaller number of flights gets done daily. Today, one of the bottlenecks of analysis for terminal and operational planners consists in realistically modeling and analyzing passenger operations, constrained by the terminal's physical design.

Air transport is one of the most influential services in Colombian economy. The country's topography and the consequent difficult access to the most remote regions make it really important. In addition, air transportation presents to Colombia, a developing country, as a gate to global economy, more convenient than other transportation options after the cost/ benefit analysis.

There has been a notable growth of air transportation system in Colombia in the last years. Between 1990 and 1999, air transportation's contribution to the Gross Domestic Product (GDP) grew to an annual average rate of 2, 7% and went down to 1, 3% in the financial crisis period between 1999 and 2002. After the crisis period, from 2003 to 2009 it grew around 4,1% similar to that of the country's economy, that tells that the demand for air transportation services is directly proportional to the Colombian economy growth [1].

Studies supporting the previous claim include: "*La infraestructura de transporte en Colombia*" by *Cardenas et al.* [2], "*El impacto del transporte aéreo en la economía colombiana y las políticas públicas*" by *Olivera et al*. [1], "*La infraestructura aeroportuaria del Caribe colombiano*" by *Otero* [3] and "*A Study of Cargo Receipt Logistics for Flower Exportation at El Dorado International Airport in Bogotá D.C.*" by *Gutierrez et al.* [4].

Research on this topic is uncommon in Latin American countries, let alone, research using modeling and simulation techniques, that and the importance of an optimal functioning air-transport infrastructure for a developing country like Colombia gives relevance to these kind of studies and encourages further research in this field.

After the implementation of the computational model was done, scenarios were constructed using the Simulation experiment from the Experiment Framework provided by *AnyLogic*. The results of the research were presented on a public lecture that reunited entrepreneurs, merchandisers, and civil authorities from the city council.

The remainder of this article is structured as follows. Section 2 presents a review of previous work on the use of modeling and simulation techniques to study and improve airport performance. Section 3 presents the agent-based and pedestrian modeling theory. Section 4 presents the research project's methodology. Section 5 demonstrates the scenarios generated and results analysis and, finally, in Section 6 conclusions are presented.

II. PREVIOUS WORK

Simulation has been used as a tool to improve performance in many fields. Regarding air transport terminal design, there are some examples like the one presented by Thomet & Mostoufi in a chapter of the book "*Transportation and Development, Innovative Best Practices*" [5] where they show the application of simulation techniques in the design of an air terminal in Curaçao. Simulation was done with a dynamic, object-oriented, pedestrian model fed by a realistic flight itinerary for a 24-hour day, with the aim of knowing the amount of passengers arriving and departing the airport along one day. Curcio presents another example in "*Passengers' flow analysis and security issues in airport terminals using modeling & simulation*" [6] where the International airport of Lamezia Terme in Calabria, Italy, is studied. The objective of the study was to analyze system performance under different scenarios through a simulation model implemented in *AnyLogic*. The passengers average wait time for reaching the gate area was the measure of system performance.

It has also been used on a more 'micro' level for addressing things like optimization of check-in points location in *"Optimizing the Airport Check-In Counter Allocation Problem"* by *Araujo & Repolho* [7], and reducing its number in *"A network model for airport common use check-in counter assignments"* by *Tang* [8].

Software tools for this kind of problems have been created, although they are still few, one of them is *GPenSIM* by *Davidrajuh & Lin* written in 2011[9] and designed to model and simulate Harstad/Narvik airport in Norway using a discrete-event system with emphasis on the flow capacity, defined as the number of passengers using the airport per time unit, and the average time required to board the plane once the passenger is inside the airport. Another one, popular within the research community, and with a more generic approach, is *SimWalk* Airport described by its authors as a *"specialized passenger simulation and analysis solution for airports that offers realistic modeling and evaluation of passenger operations"*, it allows to optimize airport design, passenger flows and terminal operations, by modeling airport objects (e.g., check-in and security control points) and integrating flight schedules and airport processes.

Another tool is *Space Syntax* by *Raford & Ragland* a pedestrian volume-modeling tool specially designed for pedestrian safety purposes [10].

Future applications of modeling and simulation techniques to improve air transport terminals' performance are endless.

Generic and little-thought construction and design guidelines for public buildings –often inadequately– are provided for the complexities of physical reality, in particular under emergency scenarios. Experience has shown that application of simplistic physical standards to densely occupied public buildings like airports, stadiums and shopping malls most of the time fails to provide the safety of crowds in an emergency situation. Moreover, the world today is facing an increasing danger from the dynamic and unpredictable threat of terrorism. One of the most effective ways to protect pedestrians in these public-places emergency scenarios lies in the attention given to their behavior within the building in these cases [11]. As stated by Smedresman, *"The increased frequency of natural and man-made disasters makes it important to assess and optimize evacuation plans. Emergency event modeling can help emergency management agencies develop effective evacuation plans that save lives."* [12].

III. AGENT BASED AND PEDESTRIAN MODELING THEORY

System modeling is a tool for solving real world problems [13]. Most of the time, solutions for these problems cannot be found by experimental means, because modifying system's components can result too expensive, dangerous, or literally impossible [14]. In these cases, the best option is to build a computational model of the real-world system. The modeling process implies a certain level of abstraction, where only the most relevant features of the system are included. A model is always less complex than the original system.

*AnyLogic* is a tool that provides three simulation methods: system dynamics, discrete event and agent-based simulation; these methods can be used individually or at the same time [12]. *AnyLogic* is one of the most popular tools in the market and has been used in many research fields and for different purposes, such as the distributed simulation of hybrid systems [15], spread of epidemics [16] and massive product consumption [17]. *AnyLogic* is the simulation tool chosen for this research project.

Agent-based modeling is a tool for the study of systems from the complex adaptive system perspective. This approach tries to explain macro phenomena as a result of micro level behavior among a heterogeneous set of interacting agents. Agent-based modeling allows for testing in a systematic way different hypotheses related to agent attributes, behavioral rules, and interaction types and their effect on macro level facts of the system [18].

Agent-based modeling allows building a system's model by identifying their objects (agents) and determining their behaviors, even if the whole system behavior, their key variables, and their dependencies are unknown. Once agent behavior is defined, agents can be created and put in an environment where they are allowed to interact. The system's global behavior is a result of lots of concurrent individual behaviors [12].

Agents determine their interactions with other agents and with their environment, on the basis of internalized social norms and mental models, internal behavioral rules and cognitive abilities, and formal and informal institutional rules among other factors [18]. In Agent-based modeling, agents do not have perfect knowledge of the system and make their decisions based on the perceptions they have on their environment; these perceptions do not have to include correct representations of reality and may vary among agents.

An agent is defined by its characteristics: activity, autonomy, and heterogeneity. Each agent has an activity; it acts according to the rules of the system and its own pre-programmed behavior. The agent's behavior can also be defined according to its features: goal direction, sensitivity, bounded reality, interactivity, mobility and adaptation [19]. Although agents may have a defined goal, and its behavior be 'ruled' by the system rules they can still make their own decisions, so they are autonomous. By the way, although each agent may begin as a member of a limited set of common templates, it can develop individuality through autonomous activity in the sense described previously [19].

The inherent complexity of agent-based modeling may make it seem like a technique with difficult or little application, and in fact, in many domains, agent-based modeling competes with traditional equation based approaches based on the identification of system variables and its evaluation through integrated sets of equations relating these variables. Both approaches simulate the system by constructing a model and executing it on a computer. The differences are in the form of the model and how it is executed [20].

In agent-based modeling, the model consists of a set of agents that encapsulate the behaviors of the various individuals that make up the system, and execution consists of emulating these behaviors. In equation-based modeling, the model is a set of equations, and execution consists of evaluating them.

Thus "simulation" is the general term that applies to both methods, which are distinguished as (agent-based) emulation and (equation-based) evaluation [20]. Understanding the relative capabilities of these two approaches is of great ethical and practical interest to system modelers and simulators. Choosing the wrong approach for a problem could lead to incorrect results and it may translate to emergency situations in the real world.

Traditional modeling methods such as discrete-event and queuing may not work well in areas with high amounts of pedestrian movement [12]. Traditional discrete-event simulation looks down on a system from above, designing process flows and creating entities to travel through the system. Agent-based modeling changes the perspective of the simulation from the high-level processes to that of the system entities, called agents. Instead of the processes evaluating and manipulating the agents, the agents themselves are able to gather information about their environment and react based on what they individually perceive.

The power of agent-based modeling lies in its ability to allow the agents to have some level of intelligence and to control their own decisions, thus resulting in behavior and outcomes that are more authentic in real-world systems dependent on individual actions. Pedestrians move in continuous space while they react to obstacles and one another [12]. Because of this unique perspective, agent-based simulation techniques are well suited for modeling pedestrian flows through an environment. The techniques are able to address some of the difficulties from which pedestrian modeling has suffered when trying to obtain complexity on a microscopic level [21].

An early example of the use of agent-based modeling to simulate pedestrian movement can be found in *STREETS*, a model developed by Schelhorn et al. in 1999 [22], under the idea of the importance of people movement for defining the 'vibrancy' of a town. The study outlined the possible uses of this kind of simulation and served as a beginning for such predictive models.

Life in cities is becoming increasingly crowded with people. Mass gatherings are more frequent in nowadays world than they were in past years, this can be attributed to the increase in world population and to air transportation becoming more cost-effective. This fact creates the need to find solutions to make those crowded places safer, and more efficient in terms of travel time. Pedestrian modeling can help to assess and optimize locations where pedestrian crowds move around [23]. Modeling allows collecting data about a given areas pedestrian density, ensure acceptable performance levels for service points with a hypothetical load, estimate how long pedestrians will stay in specific areas, and detect potential problems that interior changes such as adding or removing obstacles, or service points, may cause. Pedestrian traffic simulation plays an important role in nowadays construction, expansion, and other design-related projects for public buildings like airports, shopping centers and stadiums [12].

These studies can help architects improve building designs, facility owners review potential structural and organizational changes, engineers evaluate scenarios for improving capacity usage and civil authorities simulate possible evacuation routes in emergency scenarios. Since pedestrian flows can be complex, they require a full-blown simulation.

Pedestrians' behavior follows basic rules that have been determined by theoretical studies: they move at predetermined rates, they avoid physical obstacles such as walls, furniture and other people, and they use information about the crowds that surround them to adjust their movements (word-of-mouth effect). The results have been proven many times in field studies and customer applications [12].

In the design of a building that will have many high-traffic areas, like a supermarket, a subway station, a museum, a stadium or an airport it should be a goal to create a physical layout that minimizes travel time and ensures pedestrian flows don't interfere with each other. Simulation can test for normal, special, or peak pedestrian volume conditions, and it can also be used to understand how changes in the physical layout like the establishment of a new kiosk, new furniture or the relocation of existing items like advertising panels, flowerpots, pictures etc. will affect their operations, pedestrian travel times, and the general customer experience.

Different approaches to pedestrian modeling can be classified according to their level of abstraction, detail of description and model's time type (discrete or continuous) [23], a classification of pedestrian modeling approaches is shown in Table 1.

IV. METHODOLOGY

The aim of this project is to model the Vanguardia National Airport's passenger flow in order to obtain qualitative information that can be useful for improving airport's capacity usage, and passenger satisfaction.

Modeling and simulation are the quantitative research methods used in this project; qualitative research is done through previous work review and data collection. Data for developing the model were collected by means of observation, interviews to airport's employees and information requests made to the airport's administration. Airport's security policies limited the data collection process, as much information could not be made publicly available.

The study object is the Vanguardia National Airport (IATA: VVC, ICAO: SKVV) located in Villavicencio, Meta, Colombia. It serves domestic flights for commercial passenger, *Chárter*, cargo, and private airlines.

The airport works seven days a week, every week of the year, from 06:00 to 18:00 COT. According to airport's administration, the airport handles a total daily operation volume of about a hundred flights (this includes, all types of flights: passenger, *Chárter*, private and cargo flights) and a monthly total volume of around 3.400 flights. The number of passengers received for the year 2014 was 107.551, there were 108.121 passengers shipped that same year [24]. Only departing passengers are modeled in this project, incoming passengers are not within the scope of this study.

TABLE 1
PEDESTRIAN MODELING APPROACHES CLASSIFICATION

| Abstraction level | | |
|---|---|---|
| Microscopic models | Macroscopic models | Mesoscopic models |
| Describe each pedestrian as a unique entity with its own properties. | Determine the average pedestrian dynamics by densities, flows and velocities as functions of space and time. | These models are in between the other two, taking into account the velocity distribution. Mesoscopic models often include individual entities but model interactions between the m with common fields. |
| Description detail | | |
| Discrete-space models | | Continuous models |
| Sub-divide the environment into a lattice, and the spatial resolution of the model is limited by the cell size. | | Describe the spatial resolution down to an arbitrary level of detail. |
| Time | | |
| Discrete time | | Continuous time |
| If time is advanced only until next event occurs, system time is discrete. | | If there is no fixed time step in the model, system time is continuous. |

Villavicencio is the capital city of Meta department, and it is situated in the northern part of it, 85 km to the south of Bogotá, the country's capital.

It is considered the most-important commercial center of the Orinoquía region, and it has an approximate population of 495,200 inhabitants and a population density of 370.1 inhabitants per square kilometer. It is still a small city and its population is about one sixteenth of the Colombia's capital city, Bogotá population [25].

Lack of concise and relevant airport's operation historical data hindered the development of the model, the research team had to do the best it could with what it had. Model parameters were estimated based on data provided by the airport administration, interviews to airport and airlines employees, and in-site observation.

A. BOARDING LOGIC

Vanguardia is still a small national airport. Although there are more airlines working there, only local airlines –*Avianca* and *Satena*– have enough demand to need check-in point queues, therefore these are the only check-in points modeled.

Passengers can buy their tickets through a travel agency, call-center, web page, or at the airport facilities. Passenger check-in can be done on-line, by mobile app, or at the airport facilities. In-site check-in takes about one minute according to the airlines' employees interviewed. There are not different boarding types in the studied airlines, although there are *Chárter* flights, an exclusive type of flight that is not marketed through the usual sales channels, and that due to its characteristic is not within the scope of this study.

After checking-in, passengers must leave their luggage in the corresponding airline warehouse, where it is scanned. The airport police agents call passengers if abnormalities are found during the scanning process. The Airport has two security-check rooms; passengers go to the one working at the moment their boarding starts. Security check-in staff members check the passengers boarding pass and ticket, and made them pass through a metal detector. The entire security check-in process lasts no more than five minutes for each passenger, according to security check-in staff members.

Inside the airport there is a police office for migration control and prevention. At the moment of this research, control is for foreign passengers only. Police officer must check if the foreign passengers have criminal records after they have made their check-in, and before they go to the security check-in. Next step involves one of the two local airlines.

*Satena* (acronym for Servicio Aéreo a Territorios Nacionales) is a Colombian government owned airline that operates domestic routes, it is based in Bogotá, Colombia, and it was founded in 1962. *Satena* has direct flights from Villavicencio to Bogotá, Puerto Inírida, Puerto Carreño, and Mitú. These flights depart on different days of the week, and there is only one of each per day: Bogotá and Mitú on Mondays, Puerto Inírida on Tuesdays and Saturdays, and Puerto Carreño on Thursdays.

*Avianca* (acronym for Aerovías del continente americano S.A.) is Colombia's national airline and flag carrier, since December 5, 1919. It is headquartered in Bogotá it is the largest airline in Colombia, and the world's second oldest airline. It offers direct flights from Villavicencio to Bogotá only, and these flights depart twice every day.

Summarizing, the airport's boarding logic goes as follows: a passenger buys his ticket through a travel agency, call-center, web page, or at the airport facilities (in the day previous to the flight's date). Once the flight's date comes the passenger goes to the airport and makes his/her check-in, and goes to sit at the waiting hall until the flight's boarding starts, once that happens, the passenger goes through the security check-in, if the passenger is a foreigner he/she must have gone to the migration police office before, after the passenger goes through all this, he/she goes to the gate, shows the ticket to an airport's employee and then, boards the plane.

B. COMPUTATIONAL MODEL DEVELOPMENT

The computational model was developed using *AnyLogic* 7 University 7.2.0. Model describes the pedestrian flow at the airport by simulating the services like security checkpoints and check-in facilities and its queues, the boarding logic and the different kinds of routes each passenger could follow from the entrance to the gate, and a realistic flight schedule saved in a Microsoft Excel sheet. Finally, qualitative information is obtained through the model execution animation and the pedestrian density map included.

The model development was carried out in five phases.

In the first phase, a simplified pedestrian flow was defined. Passengers appear at the airport's entrance line and go to the gate in order to board their plane. In this path, passengers will have to stop and wait at some points. From the data given by the airport administration the research team concluded that the average passenger arrival rate is approximately 25 passengers per hour. From statistical work carried over direct observation data it was found that the average passenger comfortable walking speed lies in the interval between 0,61493 and 0,88841 meters per second with a mean of 0,75167 and a standard deviation of 0,13674 .

On the second phase, the airport's service point and waiting areas were defined, that is, the points that passengers have to stop at, previously mentioned. *Avianca* had two check-in points and *Satena*, one. The airport had two security check points, passengers go to the one with the shortest queue as both points are modeled as if they were always working; there was only one waiting area for passengers. The airport police office for migration control and prevention is modeled as an airport service as well. The delay (service) times for each service point are entered in the computational model through probability distributions (stochastic). The distribution chosen to model these delay times was the Triangular probability distribution, based on the knowledge of the minimum, maximum and a "guess" of the mean time. This limited sample data was obtained through interviews to the operating staff of the mentioned services.

On the third phase, types of passengers were differentiated. As there are no different kinds of boarding for commercial passenger flights in this airport, passengers are differentiated by their nationality. Colombian passengers get to the airport, check-in, and wait for their flight's boarding start, go through the security check-in and then go to the gate. Foreign passengers have to go to the airport's police office, to check their criminal records, before they pass through the airport's security check. The flight's schedule is defined in the fourth phase. The simulation has a duration of 12 hours following the airport's schedule, from 06:00 to 18:00 COT. The schedule implemented is based on the real destinies served by the airport, so it is a realistic schedule, but not a real one. Each flight has a departure and a destination times, this information is recorded in the spreadsheet file. Each flight has a passenger collection as well; this collection stores the list of passengers that has bought tickets for the flight. At this point, passengers have two attributes, nationality and flight.

On the last phase, the functions and events that tie the whole model together are defined, functions represent complex processes that connect the model's pieces and allow defining a simulation flow for it, events allow scheduling several similar information-dependent events happening at the same time, for this model the events were the boarding and the departure events (*EventoAbordaje* and *EventoSalida* in the model). Functions defined for this model were *configurarPasajero*, *comenzarAbordaje*, *planearAbordaje* and *configurarVuelos*.

The Simulation experiment from the *AnyLogic's Experiment Framework* was chosen as the tool for the debugging, validation and visual demonstration of the model. The computer-generated animation allowed the diagnosis of some programming errors, especially related to airport services parameters configuration.

*C.* SCENARIOS GENERATED AND RESULT ANALYSIS

The result analysis stage deals with the development of scenarios for the verification of productive bottlenecks and proposal of process improvements [26]. The flight schedule used for the scenarios generated is shown in Table 2.

TABLE 2
FLIGHT SCHEDULE

| Destination | Departure Date |
|---|---|
| Mitú – Satena | 23/01/2016 6:30 |
| Bogotá – Avianca 1 | 23/01/2016 7:00 |
| Puerto Inírida – Satena | 23/01/2016 8:30 |
| Bogotá – Satena | 23/01/2016 10:00 |
| Puerto Carreño – Satena | 23/01/2016 15:30 |
| Bogotá – Avianca 2 | 23/01/2016 16:30 |

The first scenario represents reality, which was simulated using the computational model using data obtained from interviews, specific information request to the airport administration and observations, and trying to emulate the airport service layout as close as possible.

The areas with highest pedestrian density were the waiting hall –that tends to get crowded very easily– and the area near the police office –that gets easily crowded as the number of foreign passengers grow– when there is only one officer in service. Check-in service works well as long queues are not common. When there is only one officer in service, this area gets really crowded. Around 4:00 p.m. the airport services start to work at full-capacity, all areas get crowded and the number of security check points shows insufficient as there are really long queues on both points.

Pedestrian density map and queue sizes in each service point are the main performance indicators for each scenario. Parameters like passenger arrival rate, number of check-in points, number of security check points and number of police officers in service are varied to generate the next three scenarios.

A total of four scenarios were generated. Table 3 shows a description of each scenario.

TABLE 3.
SIMULATION SCENARIOS

| Scenario | Passenger arrival rate per hour | Check in point number | Security point number | Police officer number | Annotation |
|---|---|---|---|---|---|
| 1 | 25 | 3 | 2 | 1 | Services collapse after 4 p.m. |
| 2 | 25 | 4 | 3 | 1 | Better performance. Services still collapse after 4 p.m. |
| 3 | 50 | 3 | 2 | 1 | The airport is in chaos. |
| 4 | 25 | 3 | 3 | 3 | Best performance scenario. |

For the second scenario the number of check-in and security checkpoints is increased by 1, other parameters were left equal. Performance in general, is much better than in scenario 1, the waiting areas have less pedestrian density, especially in the morning. Pedestrian traffic around 4:00 p.m. is still heavy; queues become long, but not so much as in Scenario 1. The police office is just as crowded as it was in Scenario 1.

What would happen if the number of passengers coming to the airport doubled? Would the current layout be enough? In the third scenario the passenger arrival rate is doubled, but the airport service desk layout remains untouched. By 8:00 a.m. the waiting hall is near full, but the rest of the airport remains empty. At 10:00 a.m. a first collapse occurs, security check queues get long and only empty themselves until 12:00 m. By 2:00 p.m. the waiting hall is full again. By 4:00 p.m., the airport gets really crowded, all airport service points are full and have really long queues, and pedestrian density is high in almost every point of the airport surface. The airport is in chaos.

The fourth scenario seems to be the one with the best performance; passenger flow remains nice all day long, although the waiting area gets crowded from noon onwards. Around 4:00 p.m., after an initial congestion, queues move faster than in the previous scenarios and the pedestrian density is much more evenly distributed. Despite the increase in police officers, the police office waiting area remains as the most crowded in the model.

## V. CONCLUSIONS

This research project aim was to study the pedestrian flow in the Vanguardia National Airport by means of computational modeling and simulation. By means of the simulation experiment potential effects of structural changes in the model's sources of delay –the airport's service points– in the airport's performance and found useful information for improving it.

The scenario generation done allowed finding that, first, the current airport service points layout would not be enough to face significant increases in the number of passengers; second, the security check points are the main bottlenecks in the airport's pedestrian flow, an increase in the number of these immediately improves the airport's performance. And third, the police officer for migration control and prevention is the next most important bottleneck in the flow, the process of checking the foreign passengers' criminal records can take so much time that not even tripling the number of officers is enough to avoid forming crowds at that point as the number of foreign passengers grow.

Generalization of the developments made in this study would open up possibilities for the application of these techniques in the design of functional spaces for other types of buildings with high levels of pedestrian traffic. Although is still somewhat challenging to create an airport's service point organization from scratch using simulation, a model could be used effectively to work on an already existing organization and suggest modifications for more efficient capacity usage and a better passenger flow [11].

Finally, this study enable decision makers to improve their understanding regarding pedestrian flows in public buildings in Villavicencio and offers the first and most detailed look yet at the Vanguardia National Airport processes, as well as suggestions for improving its performance. However, this should not be the concluding study for these topics, but one that inspires further research in this field.

ACKNOWLEDGMENT

The authors would like to express their gratitude to the University of the Llanos for supporting the efforts of the research team trough the study-group Advanced Simulation Concepts (ASC), to the Vanguardia National Airport administration staff, for their kind help and appreciation for the project, and finally, to IngenTIC, the IT solutions company, that helped fund their efforts.

REFERENCES

[1] M. Olivera, P. Cabrera, W. Bermúdez, and A. Hernández, "El impacto del transporte aéreo en la economía colombiana y las políticas públicas", Cuadernos de Fedesarrollo, Apr. 2011.
[2] M. Cárdenas, A. Gaviria, and M. Meléndez, "La infraestructura de transporte en Colombia," *Infraestructura, Transporte, Comunicaciones y Servicios Públicos*, Aug. 2005.
[3] A. Otero, "La infraestructura aeroportuaria del Caribe colombiano," *Documentos de trabajo sobre economía regional*, Feb. 2012.

[4] E. Gutierrez, F. Ballesteros, and J. Torres, "A study of cargo receipt logistics for flower exportation at el dorado international airport in Bogotá D.C.," in *Production Systems and Supply Chain Management in Emerging Countries: Best Practices* (G. Mejia and N. Velasco, eds.), pp. 61{79, Springer Berlin Heidelberg, 2012.

[5] M. A. Thomet and F. Mostoufi, "Simulation-Aided airport terminal design," in *Transportation and Development Innovative Best Practices 2008*, pp. 118 - 123, American Society of Civil Engineers, Apr. 2008.

[6] F. L. Duilio Curcio, *"Passengers' flow analysis and security issues in airport terminals using modeling & simulation"*.

[7] G. E. Araujo and H. M. Repolho, \Optimizing the airport Check-In counter allocation problem," *Journal of Transport Literature*, vol. 9, pp. 15-19, Dec. 2015.

[8] C.-H. Tang, "A network model for airport common use check-in counter assignments," *Journal of the Operational Research Society*, vol. 61, pp. 1607-1618, Oct. 2009.

[9] R. Davidrajuh and B. Lin, "Exploring airport trafficc capability using petri net based model,"*Expert Systems with Applications*, vol. 38, pp. 10923 - 10931, Sept. 2011.

[10] N. Raford and D. Ragland, "Space syntax: Innovative pedestrian volume modeling tool for pedestrian safety," *Transportation Research Record: Journal of the Transportation Research Board*, vol. 1878, pp. 66 - 74, Jan. 2004

[11] G. Smedresman, "Crowd simulations and evolutionary algorithms in floor plan design," May 2006.

[12] I. Grigoryev, AnyLogic in three days - *A quick course in simulation modeling*. 2015.

[13] J. Banks, J. S. Carson, B. L. Nelson, and D. M. Nicol, *Discrete-Event System Simulation (5th Edition)*. Prentice Hall, 5 ed., July 2009.

[14] *Applied Simulation and Optimization: In Logistics, Industrial and Aeronautical Practice*. Springer, 2015 ed., Apr. 2015.

[15] A. Borshchev, Y. Karpov, and V. Kharitonov, \Distributed simulation of hybrid systems with AnyLogic and HLA," *Future Generation Computer Systems*, vol. 18, pp. 829 - 839, May 2002.

[16] Eurosim, B. Zupancic, R. Karba, S. Blazic, S. S. for Simulation, Modelling, University., and F. od Electrical Engineering, "EUROSIM 2007 proceedings of the 6[th] EUROSIM congress on modeling and simulation, 9-13 September 2007, Ljubljana, Slovenia," 2007.

[17] M. Garifullin, A. Borshchev, and T. Popkov, "Using AnyLogic and agent-based approach to model consumer market".

[18] M. A. Janssen, *Agent-Based Modelling*. Arizona State University, Mar. 2005.

[19] A. Getchell, *\Agent-based modeling,"* June 2008.

[20] H. Van Dyke Parunak, R. Savit, and R. Riolo, "Agent-Based modeling vs. Equation Based modeling: A case study and users' guide," in *Multi-Agent Systems and Agent-Based Simulation* (J. a. Sichman, R. Conte, and N. Gilbert, eds.), vol. 1534 of *Lecture Notes in Computer Science*, ch. 2, pp. 10 - 25, Berlin, Heidelberg: Springer Berlin Heidelberg, 1998.

[21] M. Barker, "A survey on Agent-Based modeling of pedestrian movement," tech.rep., University of Central Florida, 2006.

[22] T. Schelhorn, D. O'Sullivan, M. Haklay, and M. Thurstain-Goodwin, "STREETS: An agent-based pedestrian model," Apr. 1999.

[23] A. Johansson and T. Kretz, "Applied pedestrian modeling," in *Agent-Based Models of Geographical Systems* (A. J. Heppenstall, A. T. Crooks, L. M. See, and M. Batty, eds.), pp. 451 - 462, Springer Netherlands, 2012.

[24] L. M. S. Rodríguez, *"Respuesta solicitud de información"*, Aeropuerto Vanguardia, Aeronáutica Civil Colombiana, Jan. 2016.

[25] DANE, "Resultados y proyecciones (2005-2020) del censo 2005," tech. rep., 2005.

[26] J. P. Lima, K. C. D. Lobato, F. Leal, and R. D. S. Lima, "Urban solid waste management by process mapping and simulation," Pesquisa Operacional, 2015.